\documentclass[journal,twoside,web]{ieee/ieeecolor}
\usepackage{ieee/generic}
\usepackage{cite}
\usepackage{amsmath,amssymb,amsfonts}
\usepackage{algorithmic}
\usepackage{rotating}
\usepackage{underscore}
\usepackage{tabularx}
\usepackage{graphicx}
\usepackage{algorithm,algorithmic}
\usepackage{hyperref}
\usepackage{textcomp}
\usepackage{booktabs}

\def\BibTeX{{\rm B\kern-.05em{\sc i\kern-.025em b}\kern-.08em
    T\kern-.1667em\lower.7ex\hbox{E}\kern-.125emX}}
\begin{document}

\title{Do Spatial Descriptors Improve Multi-DoF Finger Movement Decoding from HD~sEMG?}

\author{Ricardo Gonçalves Molinari and Leonardo Abdala Elias, \IEEEmembership{Senior Member, IEEE}
\thanks{This work was supported by the Brazilian Public Ministry of Labor (contract number 002118.2019). Leonardo A. Elias is a recipient of a Research Productivity Fellowship from CNPq (National Council for Scientific and Technological Development, proc. no. 316320/2023-4). Kinematic data were collected using the multi-user equipment Vicon Motion Capture System, funded by FAPESP (The São Paulo Research Foundation, proc. no. 2020/13293-0).}
\thanks{R. G. M. and L. A. E. are with the Neural Engineering Research Laboratory, Center for Biomedical Engineering, and the Department of Electronics and Biomedical Engineering, School of Electrical and Computer Engineering, University of Campinas, Brazil (molinari@unicamp.br; leoelias@unicamp.br).}}

\maketitle

\begin{abstract}
Restoring hand function requires simultaneous and proportional control (SPC) of multiple degrees of freedom (DoFs). This study systematically evaluated the multichannel linear descriptors-based block field method (MLD-BFM) against conventional feature extraction approaches for continuous decoding of five finger-joint DoFs, leveraging the spatial information of high-density surface electromyography (HD sEMG). Twenty-one healthy participants performed dynamic sinusoidal finger movements while HD sEMG signals were recorded from the proximal forearm. MLD-BFM extracted spatial features, including effective field strength ($\Sigma$), field-strength variation rate ($\Phi$), and spatial complexity ($\Omega$). Model performance was optimized (block size: $2\times2$; window: 0.15\,s) and compared with conventional time-domain features: root mean square (RMS) and the combination of mean absolute value and waveform length (MAV-WL), as well as dimensionality reduction approaches (PCA and NMF), using multi-output regression models. MLD-BFM consistently achieved the highest mean variance-weighted coefficient of determination ($\mathrm{R}^2_\mathrm{vw}$) across all models, with the multilayer perceptron (MLP) yielding the best result ($\mathrm{R}^2_\mathrm{vw} = 86.68\% \pm 0.33$). However, these gains were not statistically significant relative to time-domain features (RMS and MAV-WL), suggesting that dense multichannel recordings already encode spatial information through amplitude-based descriptors. Notably, MLD-BFM significantly outperformed dimensionality reduction techniques, confirming that preserving the spatial resolution of HD sEMG is critical for accurate multi-DoF finger movement regression. Among the descriptors, $\Omega$ captures the diversity of muscle sources in a way amplitude-based features cannot, regardless of channel count. Decoding accuracy was higher for the middle and ring fingers than for the thumb. These findings provide a reference for spatially informed myoelectric interfaces.
\end{abstract}


\section{Introduction}
\label{sec:introduction}
\IEEEPARstart{R}{estoring} natural and intuitive hand function in individuals with impaired hand motor function using controlled upper-limb prostheses remains a major challenge in neurorehabilitation engineering \cite{Cordella2016, Einfeldt2023}. Surface electromyography (sEMG) is the most widely adopted non-invasive interface for prosthetic control \cite{Song2023}, capturing the electrical signals generated by muscle fibers under neural command. While pattern recognition-based control (PRC) successfully classifies discrete hand gestures with high accuracy, natural and continuous volitional prosthetic control, or simultaneous and proportional control (SPC) across multiple degrees of freedom (DoFs), involves a far more complex problem \cite{Jiang2023}.

Regression-based approaches have been extensively explored to bridge this gap, aiming to map sEMG signals to continuous kinematic outputs. Seminal work by \cite{Muceli2012} demonstrated the feasibility of predicting multiple hand and wrist DoFs using multilayer perceptrons (MLP). Subsequent studies investigated a variety of models, including kernel ridge regression \cite{Hahne2014}, support vector machines \cite{Ameri2014}, k-nearest neighbors (KNN) \cite{Dewald2019}, and ridge regression \cite{Nowak2023}. These methods predominantly relied on time-domain features (e.g., root mean square - RMS, mean absolute value - MAV, waveform length - WL). Although promising, these conventional features may not fully exploit the rich spatial information provided by high-density sEMG (HD sEMG) arrays, which can limit their ability to decode the subtle and coordinated muscle activation patterns underlying dexterous hand movements.

The advent of HD sEMG has significantly advanced the study of muscular activity by enabling higher spatial resolution recordings. Unlike conventional sparse electrode arrays, HD sEMG captures detailed topographic information that researchers can use to disentangle complex motor commands \cite{Merletti2019}. By leveraging the spatial distribution of muscle activation patterns, HD sEMG allows a more precise characterization of the spatial organization and coordination of muscle activity during dynamic tasks.

The multichannel linear descriptors (MLD) framework, originally developed to describe global functional states in electroencephalography (EEG) \cite{Wackermann1996, Wackermann1999, Wackermann2007}, was recently adapted to HD sEMG to exploit its rich spatial information \cite{Peng2025}. The MLD quantifies three global spatial properties within defined regions of interest. The effective field strength ($\Sigma$) captures the overall intensity of muscle activation, the field strength variation rate ($\Phi$) reflects the speed of spatial field changes during dynamic contractions, and the measure of spatial complexity ($\Omega$) reflects the diversity of activity across different generators, serving as a quantification of the number of relevant sources \cite{Michel2009}. Building on this framework, Peng et al. \cite{Peng2025} proposed the multichannel linear descriptors-based block field method (MLD-BFM), which extracts MLD features from local electrode blocks and substantially improves pattern recognition accuracy for multiple hand and wrist movements. Although MLD-BFM has demonstrated strong performance in classification tasks, its potential to enhance continuous, simultaneous, and proportional regression of finger movements has not yet been fully explored.

In this study, we provide a systematic comparative evaluation of the MLD-BFM feature extraction method for continuous and proportional decoding of finger movements across five DoFs, benchmarking its performance against widely adopted feature sets: RMS, the combination of MAV and WL (MAV-WL), and dimensionality-reduced representations such as PCA and NMF. Our goal is to rigorously characterize the conditions under which spatially structured features offer advantages over classical approaches, and to identify which spatial descriptors contribute most distinctively to decoding performance. Importantly, while using RMS or MAV-WL across all 128 channels already encodes implicit spatial information through amplitude topography, MLD-BFM introduces a qualitatively different descriptor: spatial complexity ($\Omega$), which quantifies the diversity of active muscle sources and cannot be approximated by amplitude-based features alone, regardless of electrode density. This distinction motivates a rigorous comparative framework, providing a comprehensive reference for feature selection in the design of natural and intuitive myoelectric interfaces.

\section{Materials and Methods}
\label{sec:methods}

\subsection{Participants and Experimental Protocol}

Twenty-one healthy participants, including 10 women and 11 men, completed the experimental protocol. The age was 27.19 ± 6.55 years, body weight was 73.52 ± 16.06 kg, and height was 1.68 ± 0.11 m. One participant was left-handed. None of the participants reported neuromusculoskeletal disorders affecting the dominant hand. The protocol was reviewed and approved by the Research Ethics Committee of the University of Campinas (UNICAMP), Brazil, on 24 October 2024 (CAAE: 59961616.8.0000.5404). All participants provided informed consent before the experiments.

\begin{figure}
    \centering
    \includegraphics[width=1\linewidth]{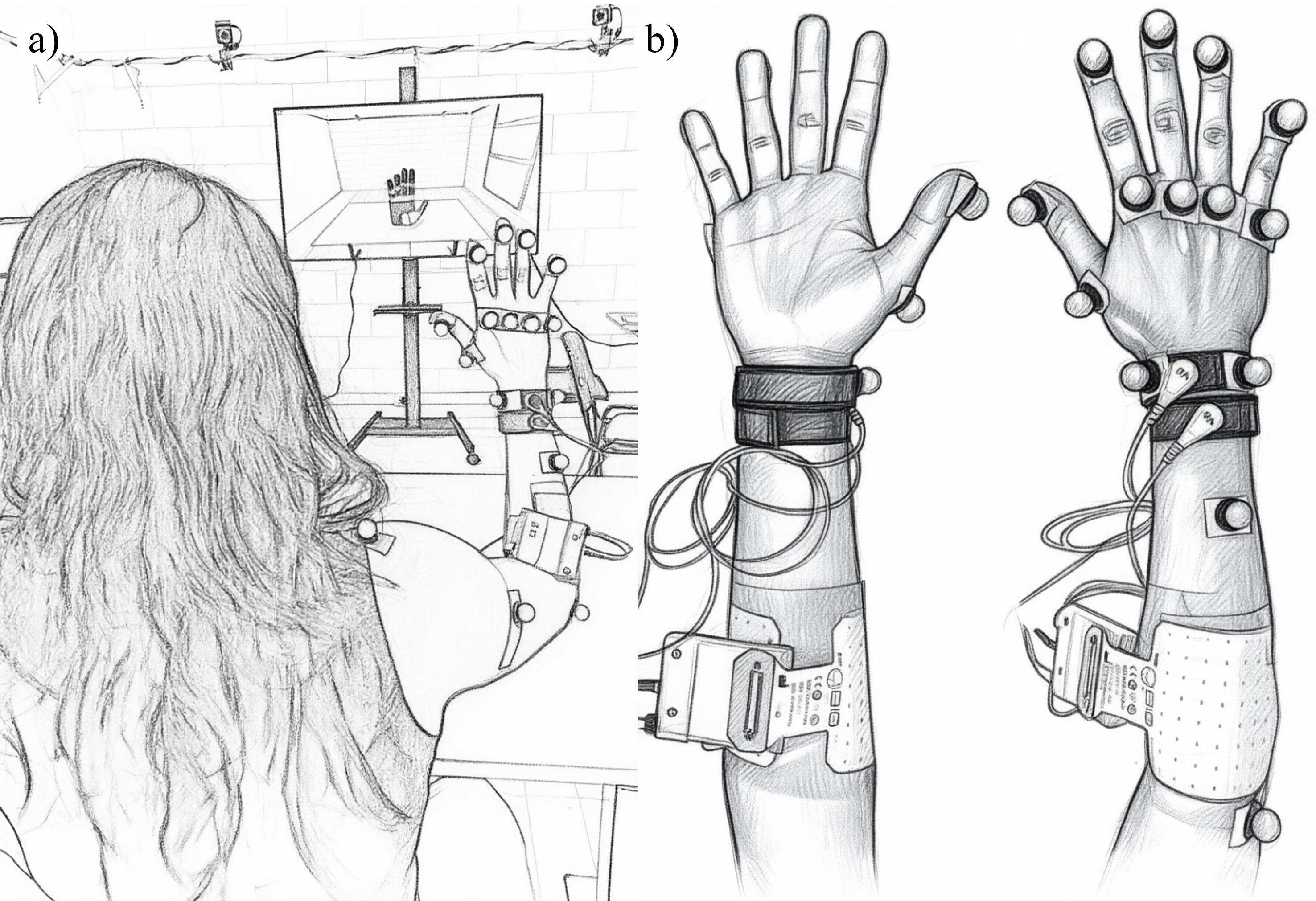}
    \caption{a) Experimental setup for the synchronous repetition of dynamic movements following a sinusoidal pattern. The finger position for all tasks was visually displayed on the screen in front of the participant. b) illustrates the positioning of the electrode arrays over the extensor digitorum communis (EDC) and flexor digitorum superficialis (FDS) muscles, as well as the reflective markers used in the motion capture system.}
    \label{fig:exp_prot}
\end{figure}

Participants followed instructions to replicate dynamic, sinusoidal finger movements displayed by a virtual hand model on a custom interface, which was displayed on a monitor placed in front of the participant while they were seated comfortably with their elbow resting on a height-adjustable table (Fig.~\ref{fig:exp_prot}A). The forearm was held in a neutral rotation position and the elbow flexed at approximately $90\,^o$. Tasks began with the fingers fully extended, followed by eight movement patterns: flexion/extension (F/E) of the index finger; F/E of the middle finger; F/E of ring and little fingers simultaneously; thumb opposition and retraction; pinch grasp and release with the index finger and thumb; pinch with the middle finger and thumb; tripod pinch with the index, middle, and thumb; and 5-finger grasping. Sinusoidal movements were displayed at 0.50 Hz and 0.75 Hz, yielding 16 trials per set. Each participant completed three sets of trials, totaling 48 tasks. The order of the tasks within each set was randomized to minimize sequence effects. In the present study, only the 0.50 Hz tasks from the first set were used, as they provided sufficient data for evaluating continuous and proportional decoding while reducing potential confounding effects related to fatigue, movement speed, and potential motor learning processes. Each task lasted 45 seconds, with at least 30 seconds of rest between tasks.

The eight movement patterns were selected to cover the finger DoF while enabling hand-function modeling within a reduced-dimensional control space for HD sEMG decoding. The set spans four primary actuation DoFs, combining isolated digit tasks with grasp-like patterns representing coordinated multi-finger synergies (pinch, tripod, and 5-finger grasp).

\subsection{Data Acquisition}

HD sEMG signals were recorded using a custom acquisition and processing platform described previously \cite{Molinari2025}. The system consisted of two 64-channel electrophysiological acquisition devices (RHD2164, Intan Technologies, USA), totaling 128 channels, with a sampling rate of 2052.52 Hz per channel. Two 64-channel electrode arrays (GR10MM0808, OT Bioelettronica, Italy) were placed over the Extensor Digitorum Communis (EDC) and the Flexor Digitorum Superficialis (FDS) \cite{Zipp1982}, following skin abrasion with an exfoliating paste and cleaning with 70\% alcohol (Fig.~\ref{fig:exp_prot}B). Reference electrodes were attached to the wrist using water-moistened conductive straps. During acquisition, signals were filtered through third-order analog Butterworth bandpass filters (10–500 Hz). The platform synchronized the start of HD sEMG recordings with the custom interface and the Vicon Vero 2.2 motion capture system (Vicon Motion Systems, UK), which operated at a frequency of 100 Hz. The kinematic data were processed using a biomechanical hand model to estimate finger joint angles.

\subsection{Block Field Method (BFM)}

The block field method (BFM) offers a structured approach to extracting spatial descriptors from HD sEMG signals by partitioning the electrode grid into local regions and computing features within each block. This approach enables the analysis of spatial activation patterns at different scales while balancing spatial resolution and computational cost.

\begin{figure*}
    \centering
    \includegraphics[width=1\linewidth]{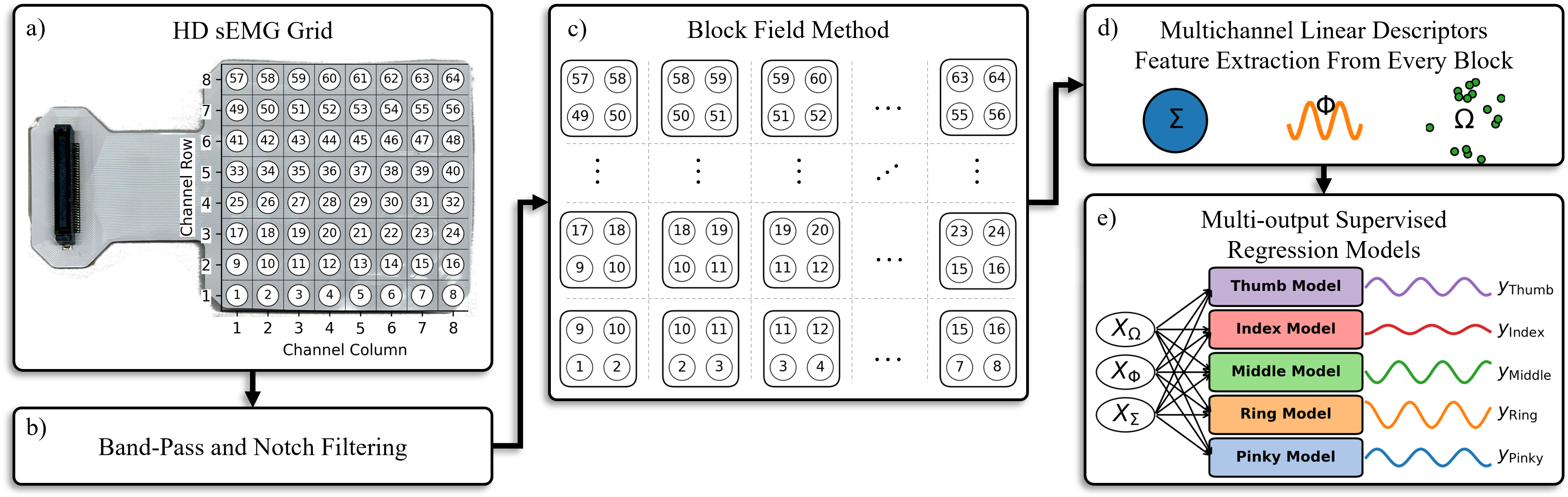}
    \caption{Schematic representation of the signal processing and feature extraction pipeline: a) electrode grid channels arranged in rows and columns; b) band-pass and notch filtering; c) block field method (BFM) example with $2 \times 2$ blocks and unit step; d) extraction of effective field strength ($\Sigma$), field strength variation rate ($\Phi$), and spatial complexity ($\Omega$); and (e) multi-output supervised regression model.}
    \label{fig:diagram_MLD-BFM}
\end{figure*}

Let $X \in \mathbb{R}^{S \times C}$ denote the filtered HD sEMG signal matrix, with $S$ samples and $C = n_\text{rows} \cdot n_\text{cols}$ channels recorded from an $8 \times 8$ electrode grid (i.e., $n_\text{rows} = n_\text{cols} = 8$). The electrode grid was segmented into spatial blocks of size $B \times B$ channels, where $B$ can range from $1$ to $8$, forming $K = B^2$ channels per block (Fig.~\ref{fig:diagram_MLD-BFM}).

The block size $B$ determines the spatial extent of each feature descriptor: smaller blocks capture finer spatial details and local variations of the HD sEMG signals, but result in a larger number of blocks $n_B$ and higher computational cost. Conversely, larger blocks provide a more global representation by aggregating signals over a wider area, reducing the total number of blocks and computational load, but potentially smoothing out finer spatial information.

In addition to the block size, the step parameter $e$ defines how the $B \times B$ blocks were distributed across the electrode grid. Specifically, $e$ determines the spatial displacement between consecutive blocks along both the row and column directions, effectively controlling the spatial sampling of the block across the electrode grid. The total number of blocks, $n_B$, was determined by the number of rows $n_{\text{rows}}$ and columns $n_{\text{cols}}$, and the block parameters $B$ and $e$ (Eq.~\ref{eq:nblocks_total}). Unlike the original implementation of the method \cite{Peng2025}, where the two lateral sides of the unfolded electrode grid were connected on the forearm, allowing blocks to slide across the lateral boundaries, in the present study, the grid was not wrapped, and therefore block placement was restricted to within the physical boundaries of each array.

\begin{equation}
n_B = \left( \left\lfloor \frac{n_\text{rows} - B}{e} \right\rfloor + 1 \right)
                 \cdot
                 \left( \left\lfloor \frac{n_\text{cols} - B}{e} \right\rfloor + 1 \right)
\label{eq:nblocks_total}
\end{equation}

In addition to the spatial configuration defined by the block size and step, temporal segmentation plays a key role in determining the resolution of the extracted features. Temporal windows of $L$ consecutive samples with overlap $O$ were defined to segment the signal in time, allowing the extraction of descriptors from short, quasi-stationary segments of the HD sEMG signals. For a given block $b$, the $w$-th temporal window restricted to the channels of the block was given by Eq.~\ref{eq:window_block}, where $\mathcal{I}_b \subset \{1, \dots, C\}$ denotes the set of indices corresponding to the $K = B^2$ channels of block $b$. Accordingly, we define $X_{w,b} \in \mathbb{R}^{L \times K}$ as the segment of the HD sEMG signals corresponding to window $w$ and block $b$.

\begin{equation}
\begin{aligned}
X_{w,b} &= X[t_w:t_w+L, \mathcal{I}_b],\\
t_w &= w (L-O), \quad w = 0, \dots, W-1
\end{aligned}
\label{eq:window_block}
\end{equation}

The window length determines the temporal extent of each segment and therefore the temporal resolution of the extracted descriptors. The overlap $O$ allows consecutive windows to share data, improving temporal continuity between segments. The number of temporal windows $W$ was determined by Eq.~\ref{eq:n_windows}. Together with the spatial block size $B$ and step $e$, the parameters $L$ and $O$ define the spatiotemporal resolution of the features extracted from the HD sEMG signals.

\begin{equation}
W = \left\lfloor \frac{N - L}{L-O} \right\rfloor + 1
\label{eq:n_windows}
\end{equation}

\subsection{Multichannel Linear Descriptors (MLD)}

The multichannel linear descriptors (MLD) framework was originally proposed for multichannel EEG analysis as a linear and physically interpretable approach to characterize global brain states \cite{Wackermann1999, Wackermann2007}. It summarizes the spatial distribution of multichannel signals using a compact set of descriptors with clear physical meaning. In this study, the framework was adapted to HD sEMG to characterize muscle activation patterns. The multichannel linear descriptors–based block field method (MLD-BFM) segments the electrode grid into spatial regions and extracts MLD features from each block \cite{Peng2025}, enabling a structured and localized representation of the spatial organization of muscular activity during different hand movements.

To formalize the feature computation, the analysis represents the HD sEMG signals in matrix form. The matrix $ X_{w,b} \in \mathbb{R}^{L \times K} $ describes the spatiotemporal segment associated with block $ b $ during temporal window $ w $, where $ K = B^2 $ denotes the number of channels within the block and $ L $ represents the number of temporal samples. Each entry $ X_{w,b}(t,k) $ contains the signal value at the $ t $-th time index ($ 1 \leq t \leq L $) and the $ k $-th channel ($ 1 \leq k \leq K $) in this segment.

The first descriptor, the \emph{effective field strength} ($\Sigma$), was originally developed to characterize the global field strength and quantifies the overall intensity of the electrical field, serving as the multidimensional analogue of Hjorth’s activity descriptor \cite{Hjorth1973}. The squared value $\Sigma^2$ corresponds to the mean integral power per channel. In HD sEMG, $\Sigma$ serves as a spatial descriptor of muscle activation patterns across different regions, analogous to the root mean square (RMS) amplitude of the signals, which provides sensitivity to contraction strength. More specifically, $\Sigma$ represents the field power within a spatial block and quantifies the intensity of muscle activity beneath the corresponding electrode region. The formal definition of $\Sigma$ was provided in Eq.~\ref{eq:sigma_new}.

\begin{equation}
\Sigma_{w,b} = \sqrt{\frac{1}{K L} \sum_{t=1}^{L} \sum_{k=1}^{K} X_{w,b}(t,k)^2 }
\label{eq:sigma_new}
\end{equation}

The second descriptor, the \emph{field strength variation rate} ($\Phi$), was originally used to describe the speed of state transitions in brain electrical activity and can be interpreted as the number of rotations per second in state space \cite{Wackermann1999}. In HD sEMG, $\Phi$ quantifies the dominant rate of change of the muscle electric field within a spatial block, serving as a generalized frequency measure (units of $s^{-1}$ or Hz) that captures the temporal dynamics of muscle activity recorded by multiple electrodes in a localized region. The mathematical definition of $\Phi$ was provided in Eq.~\ref{eq:phi_new}, where $f_s$ denotes the sampling frequency and $\Delta t = 1/f_s$.

\begin{equation}
\Phi_{w,b} = \frac{1}{2\pi} \sqrt{
\frac{\sum_{t=1}^{L-1} \sum_{k=1}^{K} \left( \frac{X_{w,b}(t+1,k) - X_{w,b}(t,k)}{\Delta t} \right)^2}
{\sum_{t=1}^{L} \sum_{k=1}^{K} X_{w,b}(t,k)^2 }
}
\label{eq:phi_new}
\end{equation}

The third descriptor, \emph{spatial complexity} ($\Omega$), was originally developed to quantify the diversity of underlying sources in EEG data and to provide a comprehensive characterization of global functional brain states \cite{Wackermann1996}. In its original formulation, $\Omega$ was derived from the normalized eigenvalues of the covariance matrix, capturing the geometrical complexity of the data in state space. The value of $\Omega$ represents the effective number of relevant sources required to describe the spatial structure of the signals \cite{Wackermann2007}. $\Omega$ ranges from 1 to $K$: it equals 1 when the data contain exactly one spatial mode, with the total variance concentrated in a single dimension, and it equals $K$ when the total variance was uniformly distributed across all $K$ modes, indicating distributed and heterogeneous activation \cite{Michel2009}. In HD sEMG, $\Omega$ serves as a spatial feature that characterizes muscle activation patterns within electrode blocks, following the same interpretation used in EEG.

The computation of $\Omega$ involves three main steps. First, the covariance matrix $C_{w,b}$ of the block signals was calculated as shown in Eq.~\ref{eq:covariance}, where the product $X_{w,b}^\top X_{w,b}$ captures the correlation structure among channels within each block. Second, the eigenvalues of this covariance matrix were normalized according to Eq.~\ref{eq:eigenvalues}, yielding $\tilde{\lambda}_i$, which define a probability distribution over spatial modes of variance. Finally, $\Omega_{w,b}$ was obtained as the entropy-like measure defined in Eq.~\ref{eq:omega_new}, quantifying the degree of spatial complexity. 

\begin{align}
C_{w,b} &= \frac{1}{L} X_{w,b}^\top X_{w,b}, \label{eq:covariance} \\[6pt]
\tilde{\lambda}_i &= \frac{\lambda_i(C_{w,b})}{\sum_{j=1}^{K} \lambda_j(C_{w,b})}, \label{eq:eigenvalues} \\[6pt]
\Omega_{w,b} &= \exp\Bigg(- \sum_{i=1}^{K} \tilde{\lambda}_i \log \tilde{\lambda}_i \Bigg) \label{eq:omega_new}
\end{align}

\subsection{Data Processing}

Feature and target matrices were generated from raw HD sEMG and finger joint angle signals, respectively, for model training and testing. The HD sEMG signals were processed using zero-phase fourth-order Butterworth bandpass filtering (10–500 Hz) and a 60 Hz notch filter with a quality factor of 30. Each task was cropped to retain data between 4 and 44~seconds, removing transient segments at the beginning and end of each motor task. Spatial features were then extracted from the filtered signals using the MLD-BFM method, and the finger joint angles were resampled to match the temporal resolution of the HD sEMG feature set.

For each task, the data were split into training and test sets with a fixed proportion of 0.5, without shuffling, to preserve the temporal structure of the signals. Specifically, the first half of each task was assigned to training and the second half to testing, ensuring that the model was always evaluated on data that followed the training period in time, thereby avoiding temporal data leakage. Feature and target matrices from all tasks were concatenated to form complete training and test datasets. The training data were then randomly permuted to improve model generalization, whereas the test data remained in sequential order to maintain temporal structure for visualization and performance evaluation. It should be noted that this protocol does not assess cross-session or cross-day generalization.

\subsection{Regression Models}

A pool of multi-output regression models was implemented to learn the mapping between HD sEMG features and finger joint angles. Each model operated within a standardized regression framework that applied input standardization and joint-output normalization, scaling all finger angles simultaneously to preserve inter-output amplitude relationships and stabilize multi-output training. Predicted trajectories were post-processed using a zero-phase fourth-order Butterworth low-pass filter with a cutoff frequency of 5 Hz.

The regression model pool was designed to provide representative coverage of the main categories of supervised learning approaches commonly used in myoelectric control research, enabling a comprehensive evaluation of the proposed MLD-BFM features under different modeling assumptions. Linear models, namely Ridge and Lasso, were included as interpretable baselines with well-established regularization properties. Ridge regression has been widely applied to simultaneous and proportional EMG decoding \cite{Nowak2023, Hahne2014}, whereas Lasso promotes sparsity, which may benefit high-dimensional feature sets through implicit feature selection.

The multilayer perceptron (MLP) was included as a representative nonlinear regressor due to its extensive use and demonstrated effectiveness in EMG-based hand kinematics estimation \cite{Muceli2012, Zhuang2019}. Ensemble methods, specifically Random Forest (RF) and Histogram-based Gradient Boosting (HGB), were selected to evaluate the contribution of tree-based nonlinear modeling \cite{Jiang2024}. These methods combine multiple learners to capture complex input–output relationships while maintaining robustness to overfitting.

Finally, k-nearest neighbors (KNN) was included as an instance-based, nonparametric regressor that makes minimal assumptions about the underlying data distribution and has shown competitive performance in dimensionality-reduced EMG feature spaces \cite{Dewald2019}. Together, these models span linear, nonlinear, ensemble, and instance-based paradigms, allowing the feature comparison to be assessed across a broad and representative set of regression strategies. Hyperparameters for each regressor were optimized using grid search with five-fold cross-validation on the training dataset, employing a variance-weighted coefficient of determination ($\mathrm{R}^2_{\mathrm{vw}}$, Eq.~\ref{eq:r2vw}) as the selection criterion (see Sec.~\ref{sec:perf_ass}). The hyperparameter search spaces were summarized in Supplementary Material Tab.~S1. Five-fold cross-validation was employed exclusively during hyperparameter optimization on the training set, and was not used for final performance evaluation.

\begin{align}
\mathrm{R}^2_{\mathrm{pred},d} &= 1 - 
\frac{\sum_{s=1}^{S} \left( y_{s,d} - \hat{y}_{s,d} \right)^2}
{\sum_{s=1}^{S} \left( y_{s,d} - \bar{y}_{d} \right)^2} \label{eq:r2pred} \\
\mathrm{R}^2_{\mathrm{vw}} &= 
\frac{\sum_{d=1}^{D} \mathrm{Var}(y_d) \cdot \mathrm{R}^2_{\mathrm{pred},d}}
{\sum_{d=1}^{D} \mathrm{Var}(y_d)} \label{eq:r2vw}
\end{align}

\subsection{Performance Assessment}
\label{sec:perf_ass}

We evaluated the multi-output regression models performance on a separate test dataset by computing the coefficient of determination ($\mathrm{R}^2_{\mathrm{pred},d}$) for each output signal $d$ (Eq.~\ref{eq:r2pred}), and then aggregated the results across all outputs using the variance-weighted formula (Eq.~\ref{eq:r2vw}), where $y_{s,d}$ and $\hat{y}_{n,d}$ denote the true and predicted values for output $d$ at sample $s$; $\bar{y}_d$ was the mean of the true values for output $d$; $S$ was the total number of samples; and $D$ was the number of output signals. The term $\mathrm{Var}(y_d)$ represents the variance of output $d$, which serves as the weighting factor in the computation of $\mathrm{R}^2_{\mathrm{vw}}$.

The coefficient of determination ($R^2$) was adopted as the base performance metric for its interpretability. To summarize performance across the five finger-joint DoFs, a variance-weighted formulation ($R^2_\mathrm{vw}$, Eq.~\ref{eq:r2vw}) was used, weighting each per-output $R^2_{\mathrm{pred},d}$ by $\mathrm{Var}(y_d)$ to emphasize outputs with larger movement amplitudes, which are more informative of the decoder's ability to track volitional commands across heterogeneous multi-output signals.

Three complementary metrics were used to evaluate the performance of each algorithm: root mean square error (RMSE, Eq.~\ref{eq:rmse}), mean absolute error (MAE, Eq.~\ref{eq:mae}), and Pearson's correlation coefficient ($r$, Eq.~\ref{eq:r}). Each metric was summarized using its variance-weighted version to account for differences in signal variability across outputs. For a given metric $M$, the global score $\bar{M}$ was computed according to Eq.~\ref{eq:variance_weighted}, where $M_d$ represents the metric value for output $d$ and $\mathrm{Var}(y_d)$ denotes the variance of the corresponding signal.

\begin{align}
\text{RMSE}_d &= \sqrt{\frac{1}{S} \sum_{s=1}^S \left( y_{s,d} - \hat{y}_{s,d} \right)^2} \label{eq:rmse} \\
\text{MAE}_d &= \frac{1}{S} \sum_{s=1}^S \left| y_{s,d} - \hat{y}_{s,d} \right| \label{eq:mae} \\
r_d &= \frac{\sum_{s=1}^S \left( y_{s,d} - \bar{y}_d \right) \left( \hat{y}_{s,d} - \bar{\hat{y}}_d \right)}
{\sqrt{\sum_{s=1}^S \left( y_{s,d} - \bar{y}_d \right)^2} 
 \sqrt{\sum_{s=1}^S \left( \hat{y}_{s,d} - \bar{\hat{y}}_d \right)^2}} \label{eq:r} \\
\bar{M}_{\mathrm{vw}} &= 
\frac{\sum_{d=1}^{D} \mathrm{Var}(y_d) \, M_d}
{\sum_{d=1}^{D} \mathrm{Var}(y_d)} \label{eq:variance_weighted}
\end{align}

\subsection{Sensitivity Analysis of Processing and Modeling Parameters}

We conducted a sensitivity analysis to quantify the impact of individual processing and modeling parameters on decoding performance. For each analysis, we systematically varied a single parameter while keeping the remaining settings fixed, allowing us to isolate its specific contribution. Across all experiments, we maintained the following control settings: a block size of $2 \times 2$, a step size of $s = 1$, a window length of 150~ms with an overlap of 50~ms, and time-windowed sequences constructed with a single-sample window ($n_{\text{win}} = 1$). These settings provided a consistent baseline for evaluating the effect of each factor.

\paragraph{Spatial-Domain parameters}  
We assessed spatial aggregation effects by varying the block size from $1 \times 1$ (single-channel) to $8 \times 8$ (full-grid). We also evaluated the impact of spatial sub-sampling by varying the step size $e$ between 1 and 6.

\paragraph{Temporal-Domain parameters}  
We investigated the effect of temporal integration by varying the window length $L$ from 100 to 500~ms in increments of 50~ms. To incorporate temporal dependencies during regression, we varied the number of consecutive samples per input sequence $n_{\text{win}}$ from 1 to 10, constructing feature vectors $X \in \mathbb{R}^{n_{\text{win}} \cdot n_{\text{feat}}}$ and target vectors $y \in \mathbb{R}^{n_{\text{win}} \cdot n_{\text{out}}}$ via a sliding window advanced by one sample. Predictions were reconstructed by concatenating the last prediction of each window to preserve the temporal structure of the original signals.

\subsection{Sequential Forward Block Selection Algorithm}

Sequential forward selection (SFS) is a greedy search algorithm that incrementally builds an optimal feature subset by maximizing a predefined performance metric, such as classification accuracy. The procedure starts with an empty set and, at each iteration, adds the single feature that provides the greatest improvement according to the objective function \cite{Aha1996}. Sequential forward block Selection (SFBS) extends the latter approach by selecting entire spatial blocks rather than individual features \cite{Peng2025}. Here, each block was represented by its corresponding MLD feature set. By evaluating blocks as unified entities, SFBS exploits the spatial organization of the data and captures the joint contribution of grouped features. 

We applied the SFBS algorithm to quantify the contribution of each spatial block to decoding performance. The set of candidate blocks was $\mathcal{U} = {1, 2, \dots, B}$, and the set of selected blocks $\mathcal{A}$ was initially empty. Each block $j$ represented a feature matrix $X_j$, and $y$ was the target vector. At each iteration, the algorithm evaluates all remaining candidate blocks by concatenating their features with those already selected (Eq.~\ref{eq:Btemp}), selects the block that maximizes the performance metric, and updates $\mathcal{A}$ accordingly.

\begin{align}
    \mathcal{B}_i &= \mathcal{A} \cup \{i\}  \label{eq:Btemp}
\end{align}

The algorithm trains a regression model using the features in $\mathcal{B}_i$ and evaluates it on the test set using the $\mathrm{R}^2_{\mathrm{vw}}$ metric to obtain a performance score $s_i$. It then selects the block $i^*$ that maximizes $s_i$ (Eq.~\ref{eq:add_block}), removes it from $\mathcal{U}$, and adds it to $\mathcal{A}$. The procedure was repeated iteratively, increasing the subset size $n{\mathcal{A}} = |\mathcal{A}|$ by one at each step until all blocks were selected.

\begin{align}
    i^* &= \arg\max_{i \in \mathcal{U}} s_i, \quad
    \mathcal{A} \gets \mathcal{A} \cup \{i^*\}, \quad
    \mathcal{U} \gets \mathcal{U} \setminus \{i^*\}  \label{eq:add_block}
\end{align}

The algorithm recorded the performance score associated with each selected block, producing a sequence of incremental scores that reflected the contribution of each block to the overall model performance. The final output was the ordered list of selected blocks and their corresponding incremental scores. Additionally, the incremental normalized $\mathrm{R}^2_{\mathrm{vw}}$ gain ($\hat{\mathrm{R}}^2_{\mathrm{vw}}$) contributed by each channel was computed and mapped onto the electrode array to reveal potential spatial patterns of channel relevance. An $\hat{\mathrm{R}}^2_{\mathrm{vw}}$ value of $1.0$ indicates the highest contribution observed for a given block in a given participant. Furthermore, the centroid of the contribution maps for both the EDC and FDS electrode arrays was calculated and reported in terms of their row and column coordinates (row, column).

\subsection{Feature Comparison}

We compared the proposed MLD-BFM representation against four alternative feature sets, selected based on their established use in the SPC literature. The chosen baselines span a spectrum from widely adopted time-domain descriptors to dimensionality reduction approaches, enabling a comprehensive and representative evaluation.

The first baseline consists of Root Mean Square (RMS) values computed from all channels. RMS is one of the most widely used amplitude-based descriptors in myoelectric control, as it provides a direct measure of the signal envelope and has been employed in regression-based SPC tasks, including ridge regression of wrist and hand kinematics \cite{Nowak2023, Hahne2014}.

The second baseline combines Mean Absolute Value (MAV) and Waveform Length (WL) from all channels (referred to as MAV-WL). This combination represents a canonical multi-feature time-domain set widely used in simultaneous and proportional control: MAV provides amplitude information analogous to RMS, while WL captures signal complexity and is sensitive to the rate of muscle activation changes. The joint use of these descriptors has been shown to yield superior regression performance compared to single-feature approaches, and constitutes a standard reference in the SPC literature \cite{Ameri2014, Hahne2014}. As such, MAV-WL serves as the primary time-domain multi-feature baseline in this study, directly addressing the question of whether spatially structured features offer advantages over conventional combined time-domain representations.

The third and fourth baselines consist of Non-Negative Matrix Factorization (NMF) and Principal Component Analysis (PCA), both applied to RMS features computed from all channels (referred to as NMF and PCA, respectively). These techniques have been explored in the HD sEMG literature as a means of capturing spatially distributed muscle activity through a compact set of components. PCA, in particular, has been applied to reduce the dimensionality of multichannel EMG envelopes while retaining the majority of signal variance \cite{Muceli2012}. NMF was included as a non-negative counterpart to PCA, better suited to the non-negative nature of EMG amplitude features. Both decomposition-based methods required selecting the number of components, which was determined individually for each participant using a plateau-based strategy, as described below.

For the decomposition-based feature sets (i.e., NMF and PCA), the optimal number of components for each participant was determined using a plateau-based strategy \cite{Cheung2005, Ballarini2021}. In this method, up to 19 components were initially extracted to construct the variance-explained ($\mathrm{R}^2_{\mathrm{var}}$, Eq.~\ref{eq:r2var}) curve as a function of the number of components. The $\mathrm{R}^2_{\mathrm{var}}$ was computed by reconstructing the original EMG RMS signals from the extracted components and comparing the reconstructed signals with the originals. The plateau was identified as the point along the $\mathrm{R}^2_{\mathrm{var}}$ curve beyond which additional components yielded only marginal gains, indicating a saturation of explained variance. Linear fits were sequentially applied to portions of the $\mathrm{R}^2_{\mathrm{var}}$ curve, and the mean squared error (MSE) of each fit was computed. The first point for which the MSE fell below a predefined threshold ($10^{-6}$ in the present study) was selected as the optimal number of components $N^*$. This procedure objectively estimated the minimal number of components that sufficiently captured the variance of the original EMG RMS feature set, balancing dimensionality reduction and information preservation.

\begin{equation}
\mathrm{R}^2_{\text{var}} = \frac{1}{C} \sum_{c=1}^{C} 
\left[
1 - \frac{\sum_{n=1}^{N} \left( x_{n,c} - \hat{x}_{n,c} \right)^2}
{\sum_{n=1}^{N} \left( x_{n,c} - \bar{x}_{c} \right)^2}
\right] \label{eq:r2var}
\end{equation}

\subsection{Statistical Analysis}

To evaluate whether the feature sets and optimization hyperparameters of the MLD-BFM yielded statistically significant differences in predictive performance, we first assessed the normality of the $\mathrm{R}^2_{\mathrm{vw}}$ distributions for each group using the D’Agostino–Pearson test. Because the data systematically violated the normality assumption, we applied non-parametric tests. Three multifactorial Kruskal–Wallis tests were conducted to examine overall differences across: (i) optimization hyperparameters and regression models; (ii) feature sets and regression models; and (iii) fingers, using the best-performing combination of feature set and regression model. 

Effect sizes were reported using Epsilon squared ($\varepsilon^2$) \cite{Tomczak2014}, categorized as negligible for $\varepsilon^2 < 1\%$, small for $1\% \leq \varepsilon^2 < 8\%$, medium for $8\% \leq \varepsilon^2 < 26\%$, and large for $\varepsilon^2 \geq 26\%$ \cite{Mangiafico2016}. \textit{post} pairwise comparisons were conducted using Dunn’s test with Bonferroni correction to adjust for multiple comparisons. Pairwise differences were visualized using a compact letter display generated with the insert–absorb algorithm and sweeping \cite{Piepho2004}, in which groups sharing at least one letter are not significantly different, whereas groups with no letters in common differ significantly. All tests adopted a significance level of $p < 0.05$. All metrics and descriptive statistics were reported as mean $\pm$ 95\% confidence interval.

\section{Results}

\subsection{Effect of Block Size}

\begin{figure}
    \centering
    \includegraphics[width=1\linewidth]{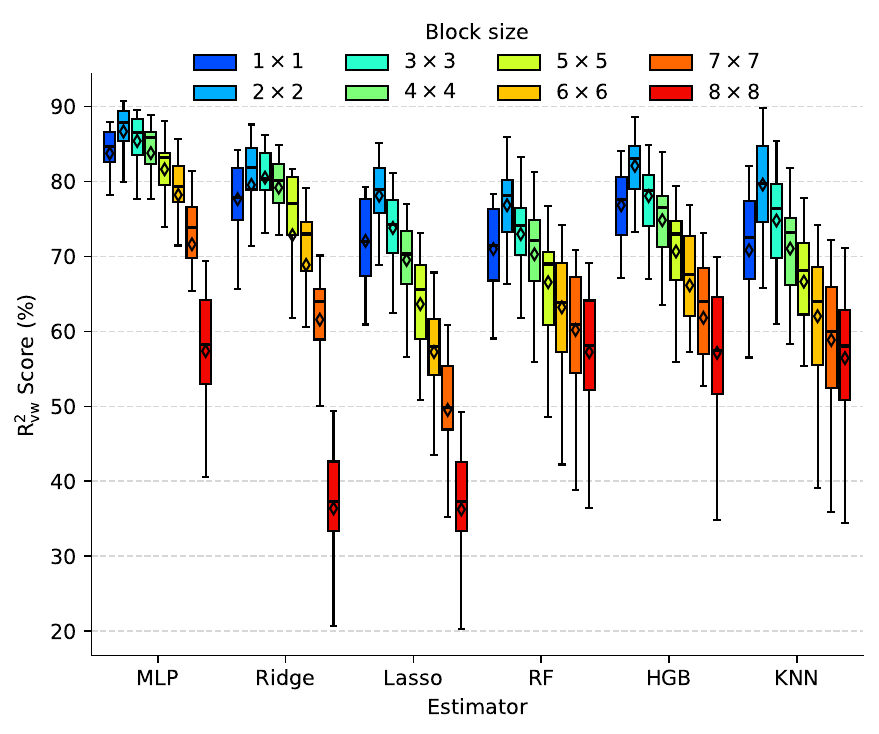}
    \caption{Distribution of $\mathrm{R}^{2}_{\mathrm{vw}}$ scores across participants, evaluating the performance of the MLD-BFM feature extraction method ($N = 21$). Results are presented for various block sizes and six regression models: Multilayer Perceptron (MLP), Ridge, Lasso, Random Forest (RF), Histogram Gradient Boost (HGB), and k-Nearest Neighbors (KNN).}
    \label{fig:block_size}
\end{figure}

Regressor performance varied systematically with block size (Fig.~\ref{fig:block_size}). The Kruskal--Wallis test revealed a large and significant main effect of block size ($p < 0.001$, $\varepsilon^2 = 48.57\%$), a medium effect of the regression model ($p < 0.001$, $\varepsilon^2 = 17.50\%$), and a large effect of their interaction ($p < 0.001$, $\varepsilon^2 = 70.50\%$) on predictive performance. Post-hoc Dunn tests confirmed that these differences were statistically significant for all regressors. For MLP, Ridge, and Lasso, block sizes 2--4 outperformed larger blocks (7--8). Ridge and Lasso also showed significant contrasts between block sizes 2--3 and intermediate sizes (6--7). Random Forest exhibited fewer but still significant differences, mainly between block size 2 and larger sizes (6--8; $p < 0.05$). HGB displayed a similar pattern, with significant differences between block sizes 1--4 and 7--8. KNN also showed significant pairwise differences between smaller (2--3) and larger (6--8) block sizes. 

Across all regressors, the highest $\mathrm{R}^2_{\mathrm{vw}}$ scores occurred at block size 2, with performance peaking for MLP ($86.68 \pm 0.33$) and remaining high for HGB ($82.09 \pm 0.38$), KNN ($79.60 \pm 0.56$), Ridge ($79.55 \pm 0.94$), and Lasso ($78.10 \pm 0.42$). Accuracy declined monotonically as block size increased, reaching substantially lower values at block size 8 across all models. Smaller blocks, particularly block size 2, provided the most effective feature representation, whereas larger blocks consistently degraded predictive performance.

\subsection{Effect of Block Step}

\begin{figure}
    \centering
    \includegraphics[width=1\linewidth]{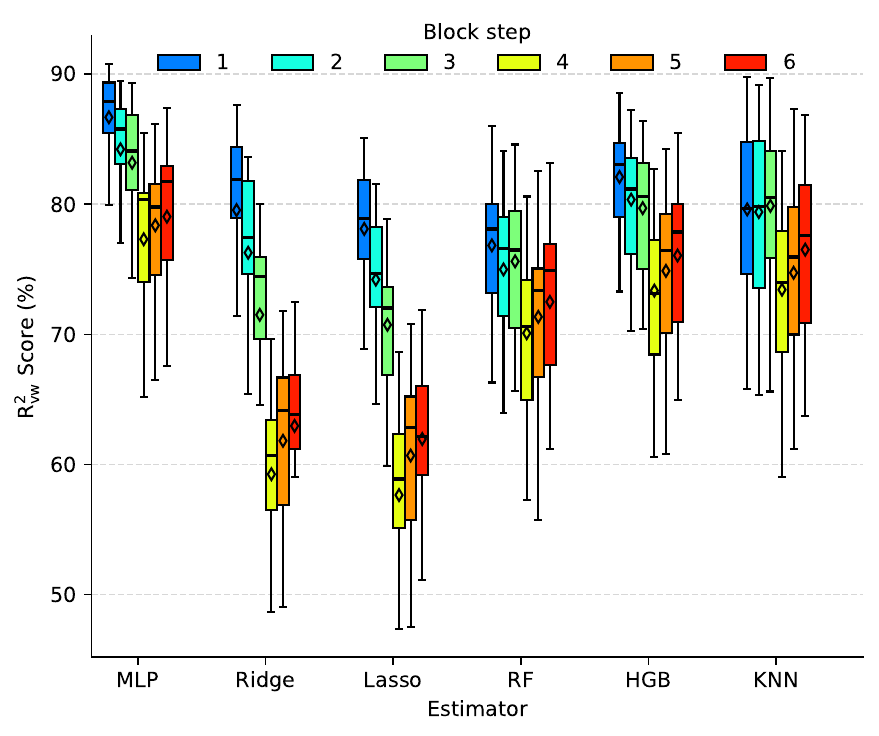}
    \caption{Distribution of $\mathrm{R}^{2}_{\mathrm{vw}}$ scores across participants, evaluating the performance of the MLD-BFM feature extraction method ($N=21$). Results are presented for various block steps and regression models, including Multilayer Perceptron (MLP), Ridge, Lasso, Random Forest (RF), Histogram Gradient Boost (HGB), and k-Nearest Neighbors (KNN).}
    \label{fig:block_step}
\end{figure}

Algorithm performance exhibited a clear dependence on block step size (Fig.~\ref{fig:block_step}). The Kruskal--Wallis test revealed a medium but significant main effect of block step size ($p < 0.001$, $\varepsilon^2 = 22.28\%$), a large effect of the regression model ($p < 0.001$, $\varepsilon^2 = 28.61\%$), and a large effect of their interaction ($p < 0.001$, $\varepsilon^2 = 55.83\%$) on predictive performance. Across all regressors, smaller block steps generally yielded higher $\mathrm{R}^2_{\mathrm{vw}}$ scores.
The highest $\mathrm{R}^2_{\mathrm{vw}}$ values were observed for MLP at block step 1, although no statistically significant differences were detected across block steps for this estimator.

\textit{Post hoc} pairwise comparisons using Dunn's test with the Bonferroni correction confirmed that these differences were statistically significant for the Ridge and Lasso models, but not for the MLP, RF, HGB, or KNN models. For Ridge, smaller block steps (1 and 2) significantly outperformed larger ones (4, 5, and 6) ($p < 0.001$), and block step 2 also showed significant advantages over block steps 5 and 6 ($p < 0.01$). Similarly, for Lasso, block steps 1 and 2 achieved significantly higher performance than block steps 4 and 5 ($p < 0.05$). These results indicate that block step size critically affects decoding performance for linear models, with small steps providing more informative feature representations, while larger steps systematically degrade predictive accuracy.

\subsection{Effects of Feature Window}

\begin{figure}
    \centering
    \includegraphics[width=1\linewidth]{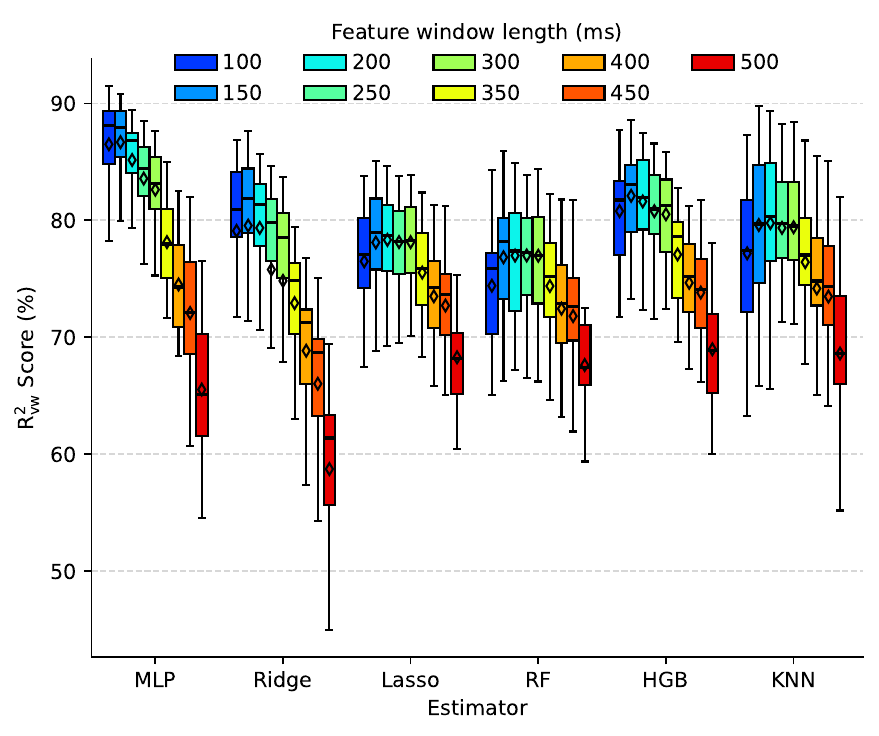}
    \caption{Distribution of $\mathrm{R}^{2}_{\mathrm{vw}}$ scores across participants, evaluating the performance of the MLD-BFM feature extraction method ($N=21$). Results are presented for various feature window lengths and regression models, including Multilayer Perceptron (MLP), Ridge, Lasso, Random Forest (RF), Histogram Gradient Boost (HGB), and k-Nearest Neighbors (KNN).}
    \label{fig:window_size}
\end{figure}

Model performance also depended systematically on the temporal window length used to compute block features (Fig.~\ref{fig:window_size}). The Kruskal--Wallis test revealed a large and significant main effect of window size ($p < 0.001$, $\varepsilon^2 = 39.51\%$), a small effect of the regression model ($p < 0.001$, $\varepsilon^2 = 7.49\%$), and a large effect of their interaction ($p < 0.001$, $\varepsilon^2 = 52.28\%$) on predictive performance. 

Post-hoc Dunn's tests confirmed significant pairwise differences for most regressors. MLP and Ridge showed markedly higher performance for intermediate windows (100–150~ms) compared to larger ones (400–500~ms). Ridge also differed from mid-range windows (250–300~ms). Lasso, HGB, and KNN exhibited similar patterns, with intermediate windows outperforming larger ones. In contrast, RF showed no significant differences, indicating a more constant performance profile.

Across all regressors, the highest $\mathrm{R}^2_{\mathrm{vw}}$ scores occurred at a window size of 150~s, with performance peaking for MLP ($86.68 \pm 0.33$) and remaining high for HGB ($82.09 \pm 0.38$), KNN ($79.60 \pm 0.56$), Ridge ($79.55 \pm 0.94$), and Lasso ($78.10 \pm 0.42$). As the window length increased beyond 150~ms, accuracy declined progressively, reaching its lowest values at 500~ms across all models. Intermediate windows, therefore, offered the best balance between temporal resolution and predictive information, whereas larger windows consistently degraded predictive performance.

\subsection{Effects of Time-Windowed Sequence Construction}

\begin{figure}
    \centering
    \includegraphics[width=1\linewidth]{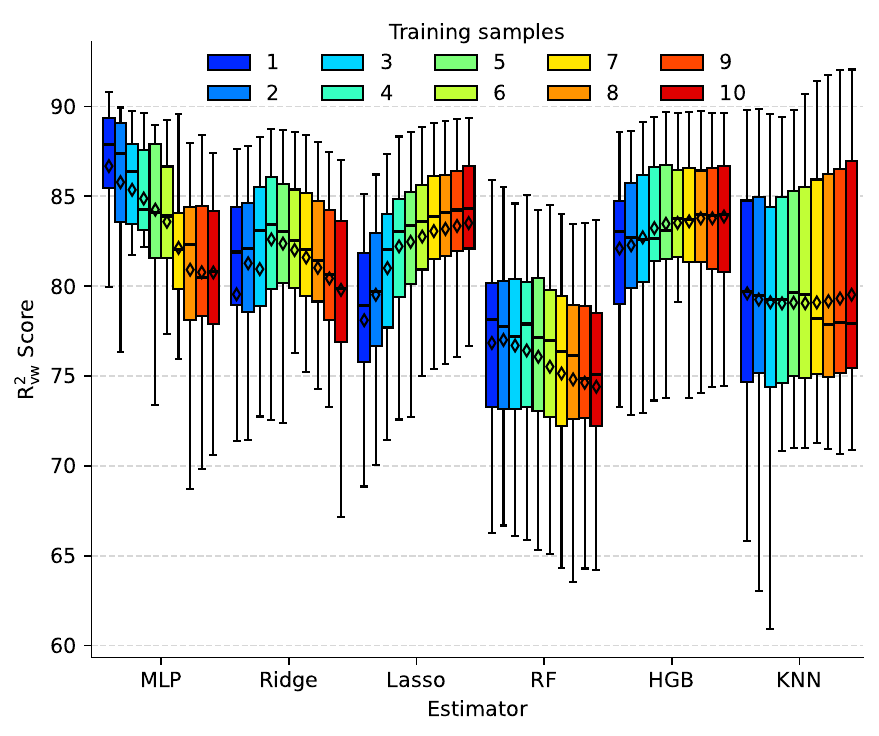}
    \caption{Distribution of $\mathrm{R}^{2}_{\mathrm{vw}}$ scores across participants, evaluating the performance of the MLD-BFM feature extraction method ($N=21$). Results are presented for various numbers of training samples and regression models, including Multilayer Perceptron (MLP), Ridge, Lasso, Random Forest (RF), Histogram Gradient Boost (HGB), and k-Nearest Neighbors (KNN).}
    \label{fig:regressor_samples}
\end{figure}

The number of training samples alone did not significantly affect model performance (Fig.~\ref{fig:regressor_samples}). The Kruskal–Wallis test revealed no main effect of sequence size ($p = 0.865$, $\varepsilon^2 = 0.37\%$), but detected medium effects of the regressor ($p < 0.001$, $\varepsilon^2 = 20.48\%$) and their interaction with sample size ($p < 0.001$, $\varepsilon^2 = 26.44\%$).

Despite the absence of a main effect of training sequence size, clear performance differences were observed across regression models. Overall, MLP consistently achieved the highest $\mathrm{R}^{2}_{\mathrm{vw}}$ values, with reference performance of $86.68 \pm 0.33$, significantly outperforming RF, KNN, and Lasso across multiple sequence sizes. Ridge, Lasso, and HGB exhibited intermediate performance levels, with representative values of $82.60 \pm 0.41$, $83.51 \pm 0.37$, and $83.88 \pm 0.34$, respectively. In contrast, RF yielded lower predictive accuracy ($77.01 \pm 0.46$), while KNN showed similarly reduced performance, with reference values near $79.52 \pm 0.81$. These results indicate that model choice played a substantially larger role than the number of training samples in determining decoding performance.

\subsection{Sequential Forward Block Selection}

\begin{figure}
    \centering
    \includegraphics[width=1\linewidth]{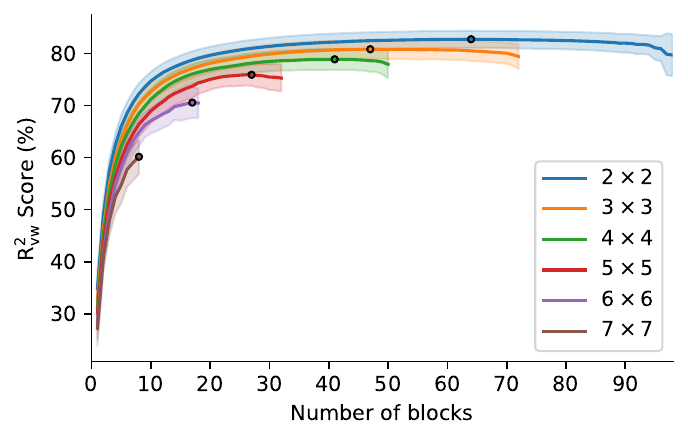}
    \caption{Performance of the Sequential Forward Block Selection (SFBS) as a function of the number of blocks and block size. The solid line represents the mean value of the population, circular markers indicate the maximum performance value for each curve, and the shaded area represents the 95\% confidence interval.} \label{fig:greed_block_combination}
\end{figure}

The Sequential Forward Block Selection (SFBS) strategy applied to the MLD-BFM feature was evaluated using the Ridge regression model due to its closed-form solution and low computational complexity, which makes it well suited for iterative feature selection procedures. Across all block sizes, $\mathrm{R}^2_{\mathrm{vw}}$ increased rapidly with the addition of the first blocks, followed by a plateau and a slight decline (Fig.~\ref{fig:greed_block_combination}). For block size 2, $\mathrm{R}^2_{\mathrm{vw}}$ rose from $34.86 \pm 3.47\%$ with the first block to a maximum of $82.74 \pm 1.64\%$ at position 64. Block sizes 3 through 7 exhibited similar rapid initial gains, with initial $\mathrm{R}^2_{\mathrm{vw}}$ values ranging from $27.23 \pm 3.38\%$ to $32.59 \pm 3.53\%$. Overall, smaller block sizes reached higher peak $\mathrm{R}^2_{\mathrm{vw}}$ values but required more sequential positions to do so, whereas larger block sizes peaked earlier but at lower $\mathrm{R}^2_{\mathrm{vw}}$ levels. The slight decline observed after the maximum likely reflects overfitting or redundancy introduced by including additional, less informative, overlapping blocks.

\begin{figure}
    \centering
    \includegraphics[width=1\linewidth]{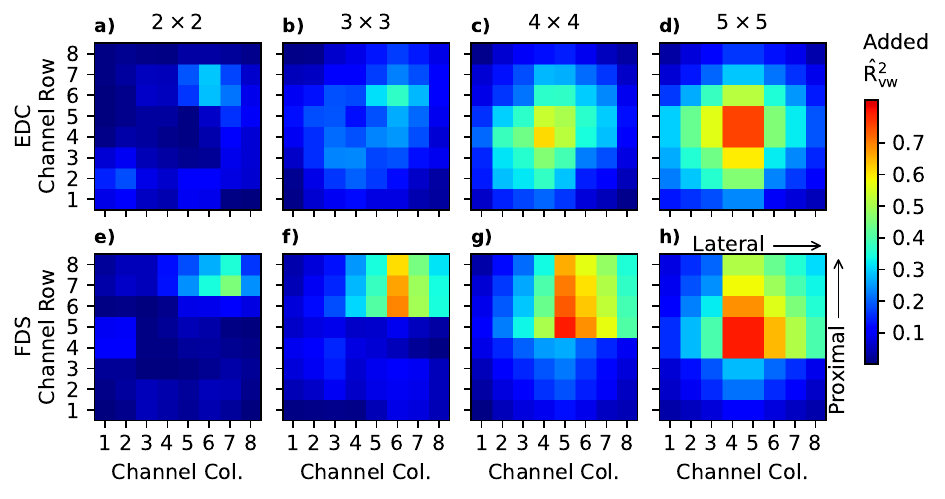}
    \caption{Heatmaps showing the average contribution of each channel in the electrode array, placed over the Extensor Digitorum Communis (EDC) and Flexor Digitorum Superficialis (FDS), to the overall performance across participants. Panels a–d) correspond to the EDC with block sizes from $2 \times 2$ to $5 \times 5$, while panels e–h) correspond to the FDS with the same block sizes.}
    \label{fig:greed_heatmap}
\end{figure}

The contribution of individual electrodes to performance, measured by the added $\hat{\mathrm{R}}^2_{\mathrm{vw}}$ across block sizes ranging from $2 \times 2$ to $5 \times 5$, is shown in Fig.~\ref{fig:greed_heatmap}. Analysis of the contribution maps revealed consistent spatial patterns across block sizes. For the $2 \times 2$ and $3 \times 3$ blocks, the highest contributions were clustered in the top-right portion of both electrode arrays. This localized pattern suggests that the linear descriptors contain highly relevant information concentrated in a specific region.

The centroid analysis revealed systematic spatial shifts as block size increased from $2 \times 2$ to $5 \times 5$ for both the EDC and FDS arrays. For the EDC array, the centroid started at $(4.34 \pm 0.56,\; 4.97 \pm 0.53)$ with block size $2 \times 2$ and progressively shifted toward the center, reaching $(4.37 \pm 0.31,\; 4.35 \pm 0.30)$ with block size $5 \times 5$. This displacement was characterized by minimal variation along the column axis, while a consistent shift along the row axis toward more distal regions was observed, with the trajectory progressing from the upper-right corner toward the center of the electrode grid. For the FDS array, the shift was more pronounced: the centroid moved from $(6.09 \pm 0.44,\; 5.26 \pm 0.43)$ for block size $2 \times 2$ to $(5.21 \pm 0.23,\; 4.99 \pm 0.27)$ for block size $5 \times 5$. This change reflects both a distal and medial shift of the added $\hat{\mathrm{R}}^2_{\mathrm{vw}}$. 

\begin{figure}
    \centering
    \includegraphics[width=1\linewidth]{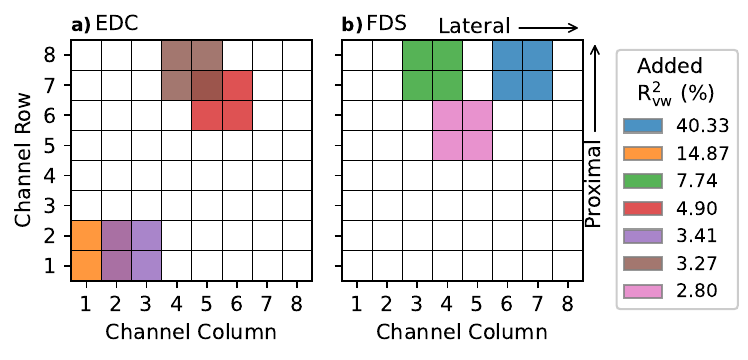}
    \caption{Representative example of the relative positions of the seven most relevant blocks for Ridge regression algorithm within the electrode arrays. (a) Array over the Extensor Digitorum Communis (EDC) muscle. (b) Array over the Flexor Digitorum Superficialis (FDS) muscle.}
    \label{fig:mld_example_map}
\end{figure}

To illustrate these spatial patterns in greater detail, Fig.~\ref{fig:mld_example_map} shows the spatial arrangement of the first seven feature blocks selected by the SFBS method using $2 \times 2$ blocks within the EDC and FDS arrays. The corresponding block feature set signals and regression model estimates across all tasks and individual fingers are provided in the Supplementary Material Fig.~S1.
 
\subsection{Regression Performance on Different Feature Sets}

\begin{figure}
    \centering
    \includegraphics[width=.8\linewidth]{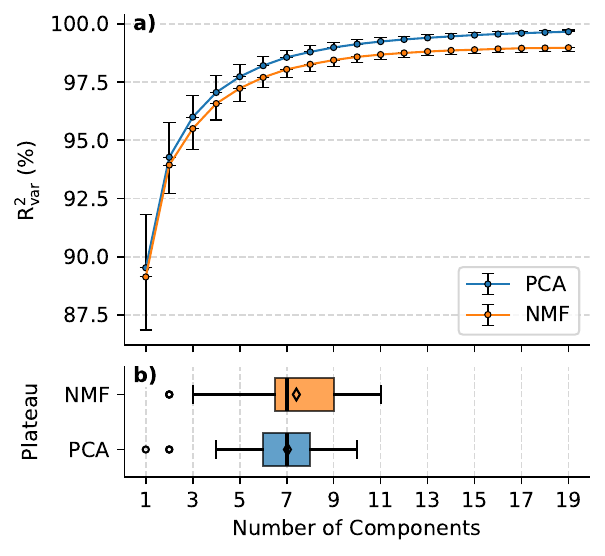}
    \caption{(a) Relationship between the explained variance of the reconstructed RMS values of HD sEMG signals and the number of components for the Non-negative Matrix Factorization (NMF) and Principal Component Analysis (PCA). Markers indicate mean values, and error bars represent the 95 \% confidence interval ($N=21$). (b) Distribution of the number of components determined by the plateau method for NMF and PCA. Diamonds denote mean values, boxes represent the interquartile range, and whiskers indicate minimum and maximum values (excluding outliers - white circles).}
    \label{fig:opn_n_comps}
\end{figure}

To evaluate the regression performance of different feature sets, we first determined the optimal number of components for PCA- and NMF-based dimensionality reduction methods. Both methods exhibited a rapid initial increase in $\mathrm{R}^2_{\mathrm{var}}$ as components were added, followed by a clear saturation trend (Fig.~\ref{fig:opn_n_comps}). The plateau values reached $98.57 \pm 0.31$ \% for PCA and $98.05 \pm 0.34$ \% for NMF, with $7.03 \pm 0.46$ \% and $7.41 \pm 0.52$ \% components, respectively. Although the optimal number of components was determined individually for each participant, these values reflect the overall trend of the participant sample as a whole. Taken together, the results show that for both techniques, approximately seven components were sufficient to capture nearly all the variance of the RMS values of HD sEMG signals, with additional components providing only marginal gains in explained variance.

\begin{figure}
    \centering
    \includegraphics[width=1\linewidth]{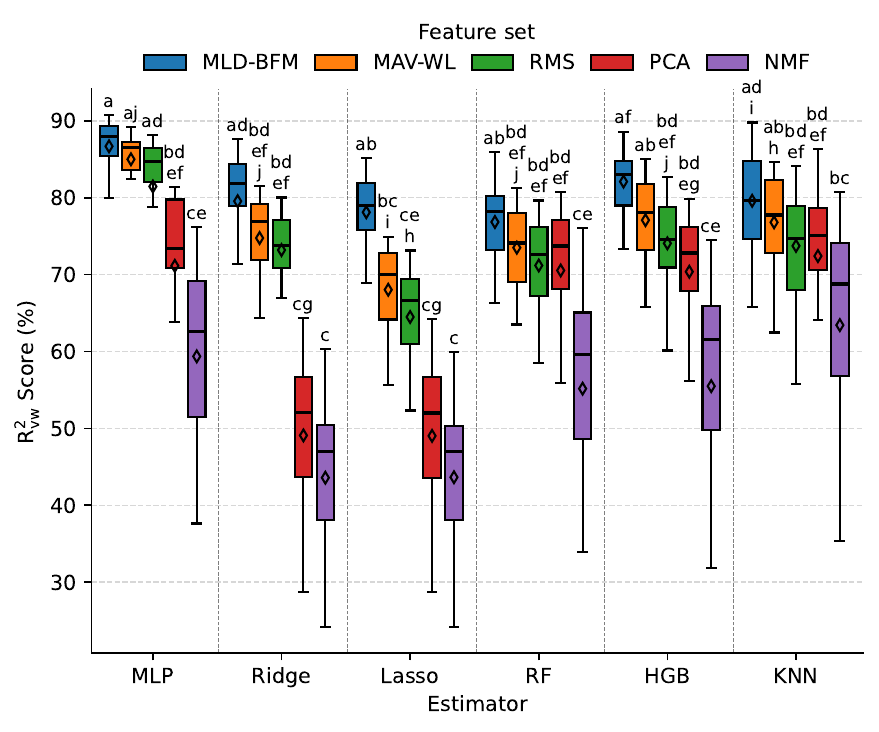}
    \caption{Distribution of the $\mathrm{R}^{2}_{\mathrm{vw}}$ score ($N=21$) as a function of the feature set (MLD-BFM, MAV-WL, RMS, PCA, and NMF) and the regression model (Multilayer Perceptron — MLP, Ridge, Lasso, Random Forest — RF, Histogram Gradient Boosting — HGB, and k-Nearest Neighbors — KNN). Diamonds denote mean values, boxes represent the interquartile range, and whiskers indicate minimum and maximum values (excluding outliers). \textit{Post hoc} comparisons (Dunn's test with Bonferroni correction, $p < 0.05$) were summarized using a compact letter display to ensure clarity. Groups that do not share a letter are significantly different.}
    \label{fig:features_comparison}
\end{figure}

The Kruskal–Wallis test revealed a significant and large effect of the feature set ($p < 0.001$, $\epsilon^2 = 43.96$~\%), a medium effect of the estimator model ($p < 0.001$, $\epsilon^2 = 15.73$~\%), and a large effect of their interaction ($p < 0.001$, $\epsilon^2 = 66.46$~\%) on $\mathrm{R}^2_{\mathrm{vw}}$. \textit{Post hoc} Dunn's tests indicated that NMF differed significantly from MLD-BFM across all regressor models. PCA also differed significantly from MLD-BFM for most regressors, except for KNN and RF, for which no significant differences were observed. Furthermore, PCA and NMF showed significant differences from MAV-WL and RMS features when combined with MLP, Ridge, and Lasso. The results are illustrated in Fig.~\ref{fig:features_comparison}, which presents each feature–regressor combination along with the corresponding \textit{post hoc} comparisons between feature sets for each regression model.

MLD-BFM achieved the highest $\mathrm{R}^2_{\mathrm{vw}}$ values across all models, with MLP reaching the best overall performance ($86.68 \pm 0.33$~\%) and HGB following closely ($82.09 \pm 0.38$~\%). KNN and Ridge yielded intermediate values ($79.60 \pm 0.56$~\% and $79.55 \pm 0.94$~\%, respectively), whereas Lasso and RF produced the lowest scores ($78.10 \pm 0.42$~\% and $76.84 \pm 0.48$~\%). Despite achieving the highest values, MLD-BFM did not differ significantly from MAV-WL and RMS features for most models. MAV-WL and RMS also showed strong predictive performance, with MLP reaching $84.99 \pm 0.38$~\% and $81.46 \pm 0.79$~\%, respectively, and HGB yielding slightly lower values ($77.08 \pm 0.52$~\% and $74.06 \pm 0.57$~\%). PCA and NMF generally underperformed, with their best results obtained with KNN ($72.41 \pm 0.89$~\% for PCA and $63.43 \pm 1.44$~\% for NMF). Table~\ref{tab:regression_results} summarizes the best-performing regressor for each feature set.

\begin{table}
\centering
\caption{Results obtained for each feature set using its best-performing regressor model.}
\label{tab:regression_results}
\scriptsize
\setlength{\tabcolsep}{3pt}
\begin{tabularx}{\columnwidth}{X c c c c}
\toprule
Feature (estimator) & $\mathrm{R}^2_{\mathrm{vw}}$ Score (\%) & RMSE (${}^o$) & MAE (${}^o$) & Pearson $r$ \\
\midrule
MLD-BFM (MLP) & 86.68 $\pm$ 0.34 & 10.98 $\pm$ 0.17 & 7.99 $\pm$ 0.12 & 0.93 $\pm$ 0.00 \\
MAV-WL (MLP) & 84.99 $\pm$ 0.39 & 11.72 $\pm$ 0.19 & 8.63 $\pm$ 0.14 & 0.92 $\pm$ 0.00 \\
RMS (MLP) & 81.46 $\pm$ 0.81 & 12.86 $\pm$ 0.26 & 9.32 $\pm$ 0.18 & 0.90 $\pm$ 0.00 \\
PCA (KNN) & 72.41 $\pm$ 0.92 & 15.76 $\pm$ 0.30 & 11.01 $\pm$ 0.25 & 0.84 $\pm$ 0.01 \\
NMF (KNN) & 63.43 $\pm$ 1.47 & 18.10 $\pm$ 0.40 & 12.62 $\pm$ 0.31 & 0.79 $\pm$ 0.01 \\
\bottomrule
\\[-0.5em]
\multicolumn{5}{l}{\footnotesize\parbox{\linewidth}{\textit{Acronyms}: 
MLD-BFM -- Multichannel Linear Descriptors-based Block Field Method; 
MAV-WL -- Mean Absolut Value and Waveform Length; 
RMS -- Root Mean Square; 
PCA -- Principal Component Analysis; 
NMF -- Non-negative Matrix Factorization; 
MLP -- Multilayer Perceptron; 
KNN -- k-Nearest Neighbors;
$\mathrm{R}^2_{\mathrm{vw}}$ -- Variance-weighted coefficient of determination; 
RMSE -- Root Mean Square Error; 
MAE -- Mean Absolute Error.}}
\end{tabularx}
\end{table}

\begin{figure}
    \centering
    \includegraphics[width=1\linewidth]{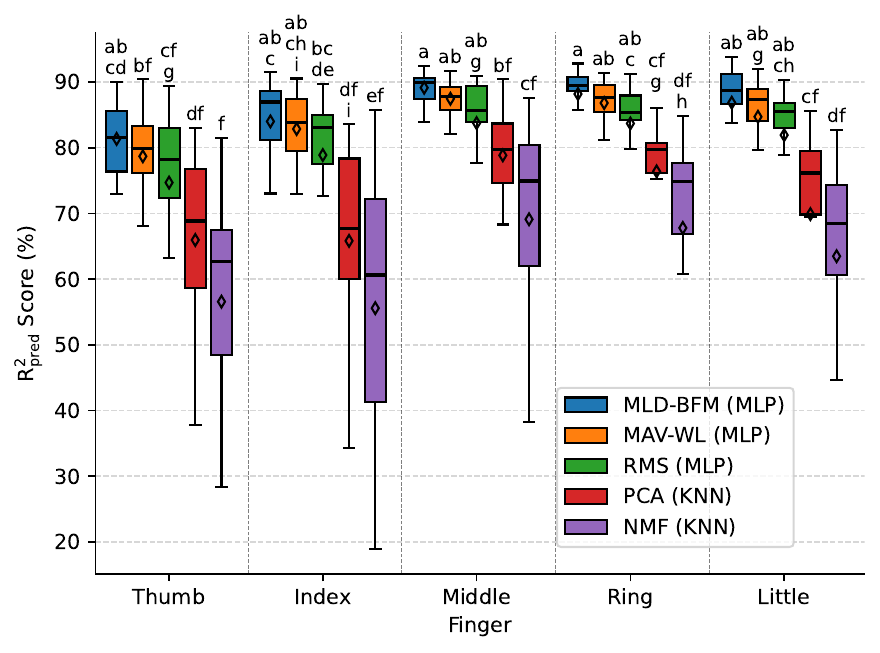}
    \caption{Distribution of the $\mathrm{R}^{2}_{\mathrm{pred}}$ score for predicting the angular position of each finger (thumb, index, middle, ring, and little) using the best-performing model for each feature set. Diamonds denote mean values, boxes represent the interquartile range, and whiskers indicate minimum and maximum values (excluding outliers). \textit{Post hoc} comparisons (Dunn's test with Bonferroni correction, $p < 0.05$) were summarized using a compact letter display to ensure clarity. Groups that do not share a letter are significantly different.}
    \label{fig:finger_comparison}
\end{figure}

The results for individual fingers across feature sets and regression models are presented in Fig.~\ref{fig:finger_comparison}. The Kruskal–Wallis test revealed a large effect of the feature–regressor factor ($p < 0.001$, $\varepsilon^2 = 42.88$~\%), a medium effect of the finger factor ($p < 0.001$, $\varepsilon^2 = 11.24$~\%), and a large effect of their interaction ($p < 0.001$, $\varepsilon^2 = 54.87$~\%). The results indicate that most of the variance in decoding accuracy stems from the choice of feature set and regression model, that performance varies to a moderate extent across fingers, and that the interaction between feature–regressor combinations and specific accounts for a large proportion of the variance in performance.

Dunn's \textit{post hoc} tests showed that MLD-BFM with MLP consistently outperformed NMF with KNN across all fingers ($p < 0.01$) and differed significantly from PCA with KNN for most fingers, except for the thumb. No significant differences emerged between MLD-BFM (MLP), MAV-WL (MLP) or RMS (MLP), indicating statistically comparable performance among these three approaches within each finger.

MLD-BFM with MLP yielded the highest $\mathrm{R}^2_{\mathrm{pred}}$ values for all fingers, ranging from $81.34 \pm 0.53$~\% for the thumb to $89.12 \pm 0.21$~\% for the middle finger. MAV-WL (MLP) achieved slightly lower results ($78.69 \pm 0.72$~\% for the thumb and $87.45 \pm 0.30$~\% for the middle finger), confirming that simple time-domain features retain substantial predictive power. RMS features followed a similar pattern but showed a more pronounced drop for the thumb ($74.69 \pm 1.21$~\%), while still reaching $83.74 \pm 1.03$~\% for the middle finger. In contrast, dimensionality reduction approaches underperformed: PCA with KNN achieved moderate accuracy for the middle ($78.84 \pm 0.67$~\%) and ring ($76.39 \pm 0.90$~\%) fingers but dropped to approximately 66\% for the thumb and index. NMF with KNN consistently yielded the lowest performance, with values as low as $55.58 \pm 1.85$~\% for the index finger.

\section{Discussion}

The present study aimed to systematically evaluate the MLD-BFM for the simultaneous and proportional decoding of five DoFs of finger joint movements, leveraging the rich spatial information provided by HD sEMG. The findings indicate that MLD-BFM consistently achieved the highest performance values across all tested regression models, with the MLP model reaching the overall best performance. While conventional time-domain features also demonstrated strong predictive capabilities and were statistically comparable to MLD-BFM for most models, dimensionality reduction techniques such as PCA and NMF yielded substantially lower accuracy comparatively. Overall, the results highlight the importance of preserving the spatial richness of HD sEMG for SPC, providing evidence that spatially structured features enhance decoding accuracy and can contribute to the development of more natural and responsive real-time myoelectric interfaces.

\subsection{Influence of Spatial Parameters on MLD-BFM Decoding Performance}

The parameter analysis of MLD-BFM demonstrated that block size constitutes a critical determinant of prediction accuracy, with the $2 \times 2$ configuration yielding the highest performance. The result is consistent with previous findings obtained using classification-based algorithms \cite{Peng2025}. The present study extends this observation to continuous multi-DoF regression, providing early evidence of the suitability of MLD-BFM features for regression models. The superior performance obtained with small spatial blocks ($2 \times 2$) reflects an optimal trade-off between capturing local spatial dependencies and preserving signal complexity across channels, thereby maximizing feature expressiveness, whereas larger blocks may attenuate descriptor discriminability.

The findings indicated that smaller block steps consistently outperformed larger ones. This dependency was particularly critical for linear regression models (Ridge and Lasso). The block step determines the density and distribution of the blocks, and smaller steps maximize overlap, preserving continuity and spatial interactions between neighboring regions. As a result, they provide more informative feature representations and finer spatial resolution. Conversely, increasing the block step can be leveraged to reduce computational complexity and accelerate model processing, creating a relevant trade-off between accuracy and computational efficiency—especially in real-time applications with hardware constraints.

In the context of the SFBS technique, as the block size increased, the centroid of the added $\hat{\mathrm{R}}^2_{\mathrm{vw}}$ progressively converged toward the center of both electrode grids. This trend reflects the greater spatial overlap of larger blocks with central regions and their correspondingly higher contribution to algorithm performance. Although systematic medial and distal shifts in the centroid were observed with increasing block size, the spatial locations of the most informative regions remained largely consistent within each electrode grid, particularly in the proximal–lateral areas of both the EDC and FDS grids. These findings suggest that the underlying activation patterns were preserved across block sizes and participants, and that the regions contributing most strongly to decoding performance remain robust to variations in block size.

It is worth noting that the spatial sliding-window structure of BFM bears a conceptual resemblance to the local receptive fields used in convolutional neural networks (CNNs), where a kernel of fixed size is applied across a spatial input to extract local features \cite{araujo2019computing}. However, a fundamental distinction separates the two approaches. In CNNs, the convolutional kernels are learned from data through backpropagation and lack direct physical interpretation. In contrast, BFM applies fixed, mathematically defined operations within each spatial block, yielding descriptors, $\Sigma$, $\Phi$, and $\Omega$, with explicit physical meaning rooted in field theory and information theory. This makes MLD-BFM a transparent and interpretable feature extraction framework, whose outputs can be directly related to the intensity, dynamics, and source complexity of muscle activation, independently of training data. This interpretability may be particularly valuable in clinical and rehabilitation contexts, where understanding the physiological basis of decoder inputs is as important as predictive accuracy.

\subsection{Influence of Temporal Parameters on MLD-BFM Decoding Performance}

While previous studies employing classification models reported improved performance with increasing temporal window lengths \cite{Peng2025}, our results indicate an optimal window duration of approximately 150~ms, yielding the highest $\mathrm{R}^2_{\mathrm{vw}}$ scores across all tested regressors. This discrepancy likely reflects differences in experimental design and study objectives. In \cite{Peng2025}, participants performed discrete movements held for 4~s for the classification of static contractions, whereas the present work focuses on the continuous regression of dynamic sinusoidal finger movements. The optimality of the 150 ms window likely reflects a balance between capturing the temporal dynamics of muscle activation during continuous finger movements and avoiding excessive temporal smoothing that may obscure relevant signal variations.

Although natural hand movements are typically aperiodic and task-dependent, sinusoidal or continuous tracking paradigms are widely used in the evaluation of myoelectric regression algorithms because they provide controlled coverage of the full range of motion while enabling quantitative comparisons across decoding strategies \cite{Jiang2012, Muceli2012, Hahne2014, Simpetru2023}. Importantly, the models in this work were trained exclusively from HD sEMG inputs without autoregressive kinematic information. In addition, the input–output training pairs were randomly shuffled, preventing the models from exploiting the temporal or periodic structure of the sinusoidal trajectory. Consequently, the regression models learn the mapping between muscle activation patterns and joint kinematics rather than the phase of the movement itself.

Beyond window length, movement speed has also been identified as a critical factor influencing algorithm performance, as fast and forceful movements generate large, distinctive EMG patterns, whereas slow movements produce subtler and noisier variations in myoelectric signals \cite{Tully2025}. Indeed, previous studies have demonstrated a robust correlation between increased movement speed and the corresponding rise in myoelectric activity \cite{Porta2024}. Given that most decoders are trained on homogeneous single-speed datasets, and that limited variability in training speeds can impair generalization and degrade decoder performance, future studies should investigate whether training on heterogeneous tasks comprising both slow and fast movements alters the optimal temporal window duration.

\subsection{Influence of Feature Extraction and Regression Methods on Decoding Performance}

The feature extraction method strongly influenced the performance of the regression models. MLD-BFM achieved the highest average performance ($\mathrm{R}^{2}_{\mathrm{vw}}$) across all tested regressors, although its advantage over conventional time-domain features was not statistically significant, so that these classical amplitude-related features may remain viable alternatives when computational efficiency is a priority. Among the evaluated models, the MLP performed best when combined with MLD-BFM, likely due to its ability to efficiently capture complex nonlinear relationships. This result aligns with the widespread use of MLP as a nonlinear regressor in myoelectric control research \cite{Nielsen2011, Muceli2012, Jiang2012, Hahne2014, Zhuang2019}. However, this benefit comes at the expense of substantially longer training (see Supplementary Material Tab.~S2) times compared to linear models such as Ridge and Lasso \cite{Hahne2014}.

In contrast, the dimensionality reduction techniques explored here -- principal component analysis (PCA) and non-negative matrix factorization (NMF) -- consistently produced lower accuracy. This outcome supports previous findings \cite{Muceli2012}, suggesting that reducing the spatial dimensionality of HD sEMG signals can impair continuous decoding performance. Interestingly, instance-based regressors like KNN partially mitigated the information loss in PCA and NMF, suggesting that local neighborhood-based methods can compensate to some extent for reduced feature sets \cite{Roughan2012}.

These observations also point to an important direction for future investigation. The present comparison evaluated all feature sets under identical, controlled conditions, where amplitude-based descriptors applied across 128 channels already encode substantial implicit spatial information. Under such conditions, the distinctive contribution of spatial complexity~($\Omega$), which quantifies the diversity of active muscle sources independently of signal amplitude, may be underestimated. Future work should investigate the isolated contribution of $\Omega$ relative to amplitude-based spatial descriptors, as well as the behavior of MLD-BFM under conditions where spatial structure becomes more critical, such as inter-session variability, where a richer spatial characterization may reveal advantages not captured by accuracy metrics alone. Complementary evidence regarding robustness to electrode displacement and muscle fatigue is further discussed in Section~\ref{sec:limitations}.

\subsection{Influence of Finger on Prediction Performance}

The finger-specific analysis revealed additional insights. The middle, ring, and little fingers consistently yielded the highest predictive accuracy across feature sets and regressors, whereas the thumb exhibited the lowest predictive accuracy. A comparable pattern emerged in studies employing convolutional neural network–based models (TF2AngleNet), where the ring and little fingers achieved the best performance and thumb extension the poorest \cite{Jiang2025}. The reduced accuracy was attributed to data imbalance, with thumb movements underrepresented. In contrast, the present work included five thumb-related tasks out of eight, providing a broader representation. A similar trend was observed in a study using biomechanical models (OpenSim) driven by EMG signals, in which the thumb showed the lowest correlation with kinematic data, whereas the middle finger exhibited the highest \cite{Blana2020}.

A plausible explanation for these findings involves both anatomical and physiological factors. Electrodes are typically positioned closer to the muscle groups that predominantly control the middle and ring fingers, resulting in signals with a higher signal-to-noise ratio and greater redundancy across channels. In contrast, thumb movements rely on more distributed and anatomically complex muscle activations, some of which may fall outside the optimal detection range of the electrode array, leading to reduced predictive accuracy \cite{Blana2020}. Additionally, previous work demonstrated that the middle finger exhibits a more localized and concentrated muscle activation compared to other fingers, with its activation centroid remaining relatively stable across different forearm rotation angles \cite{Rubin2022}. This spatial consistency likely contributes to the robustness of EMG-based prediction methods for this finger. Conversely, thumb movements rely on anatomically complex and spatially distributed muscle activations that result from the coordinated combination of flexion, adduction, and pronation at the carpometacarpal joint. These movements are generated by four extrinsic and five intrinsic muscles of the thenar group, some of which may fall outside the optimal detection range of the electrode array, resulting in reduced predictive accuracy \cite{Nanayakkara2017, Blana2020}.

\subsection{Limitations}
\label{sec:limitations}

Despite the comprehensive evaluation conducted in the present study, some limitations should be acknowledged. First, the experiments were performed with a limited number of healthy participants, which may constrain the wider applicability of the findings to clinical populations or individuals with altered or reduced muscle activation patterns \cite{Daley2012, Kristoffersen2020}. Moreover, the predictive performance of the proposed approach was evaluated exclusively under controlled laboratory conditions, using sinusoidal finger movements at a fixed frequency, and performed under a single arm posture and contraction intensity. Such conditions may not fully reflect the variability and complexity of natural motor behaviors encountered in real-world scenarios \cite{Vujaklija2017, Campbell2020, Rubin2022, Tully2025}. Previous work has demonstrated that forearm rotation angle significantly affects the spatial distribution of HD sEMG signals by altering the relative position of the electrode array with respect to the underlying musculature, which can substantially impact decoding performance 
\cite{Rubin2022, Campbell2020}.

Second, sEMG signals are inherently susceptible to multiple sources of noise and variability, particularly during prolonged muscle activation. Factors such as perspiration-induced changes in skin impedance \cite{Abdoli-Eramaki2012}, muscle fatigue, and motor learning effects \cite{Kyranou2018} can alter signal characteristics over time, potentially requiring periodic updates or adaptive strategies to maintain long-term robustness of the algorithms. In addition, manual electrode placement cannot fully eliminate positioning variability, and when displacement occurs, predictive performance can deteriorate substantially due to changes in recorded muscle activity that were not represented during model training \cite{Kanoga2020, Li2021}.

\section{Conclusion}

This study provides a systematic evaluation of MLD-BFM for SPC of five finger-joint DoFs using HD sEMG. The central question addressed was whether spatially structured features offer measurable advantages over conventional time-domain approaches for continuous multi-DoF regression. The results reveal an important nuance: applying amplitude-based features such as RMS or MAV-WL across all 128 channels already encodes substantial spatial information implicitly through the topographic distribution of activation across the electrode grid. This explains why the performance gap between MLD-BFM and time-domain baselines, while consistently favoring MLD-BFM in mean accuracy, did not reach statistical significance. In this context, the most distinctive contribution of MLD-BFM lies in the spatial complexity descriptor ($\Omega$), which captures the diversity of muscle sources, a property that amplitude-based features cannot represent regardless of channel count.

MLD-BFM significantly outperformed dimensionality reduction techniques (PCA and NMF), confirming that compressing the spatial information of HD sEMG into a small number of components is detrimental to continuous regression accuracy. The optimal configuration identified was small block sizes ($2\times2$) with short temporal windows (150 ms), providing practical guidance for feature extraction in real-time myoelectric interfaces.

\section*{Acknowledgments}

We would like to express our gratitude to Valeria Avilés Carrillo and Guilherme A. G. De Villa for their assistance in data collection.



\bibliographystyle{ieeetr}
\bibliography{references}

\end{document}


\setcounter{figure}{0}
\setcounter{table}{0}
\renewcommand{\thefigure}{S\arabic{figure}}
\renewcommand{\thetable}{S\arabic{table}}
\maketitle

\begin{table}
\centering
\caption{Regression models and hyperparameter ranges optimized through cross-validation grid-search.}
\begin{tabularx}{\linewidth}{l>{\raggedright\arraybackslash}X}
\toprule
Model & Hyperparameters (default parameter or range) \\
\midrule
MLP & Hidden layer sizes: \{10, 15, 20\}, activation: 'relu', learning rate: \{0.01, 0.1\}, max iterations: 200, early stopping with patience: 20 \\
Ridge & Regularization strength: $\alpha \in \{0.001, 0.01, 0.1, 1.0, 10.0\}$ \\
Lasso & Regularization strength: $\alpha \in \{0.01, 0.1, 1.0, 10.0\}$, max iterations: 10000, tolerance: 1e-3 \\
RF & Number of trees: \{25, 50\}, maximum depth: \{10, 20\}, maximum features: \{'sqrt', 'log2'\}, bootstrap=True, max samples: 0.5 \\
HGB & Maximum iterations: 50, learning rate: \{0.01, 0.1\}, maximum depth: \{3, 5\}, max features: 0.8, early stopping with patience: 20 \\
KNN & Number of neighbors: \{10, 30, 50\}, weights: \{'uniform', 'distance'\} \\
\bottomrule
\\[-0.5em]
\multicolumn{2}{l}{\footnotesize\parbox{\linewidth}{\textit{Acronyms}: MLP -- Multilayer Perceptron; RF -- Random Forest; HGB -- Histogram-Based Gradient Boosting; KNN -- k-Nearest Neighbors.}}
\end{tabularx}
\label{tab:regressors}
\end{table}

\begin{sidewaysfigure*}
    \centering
    \includegraphics[width=\textheight]{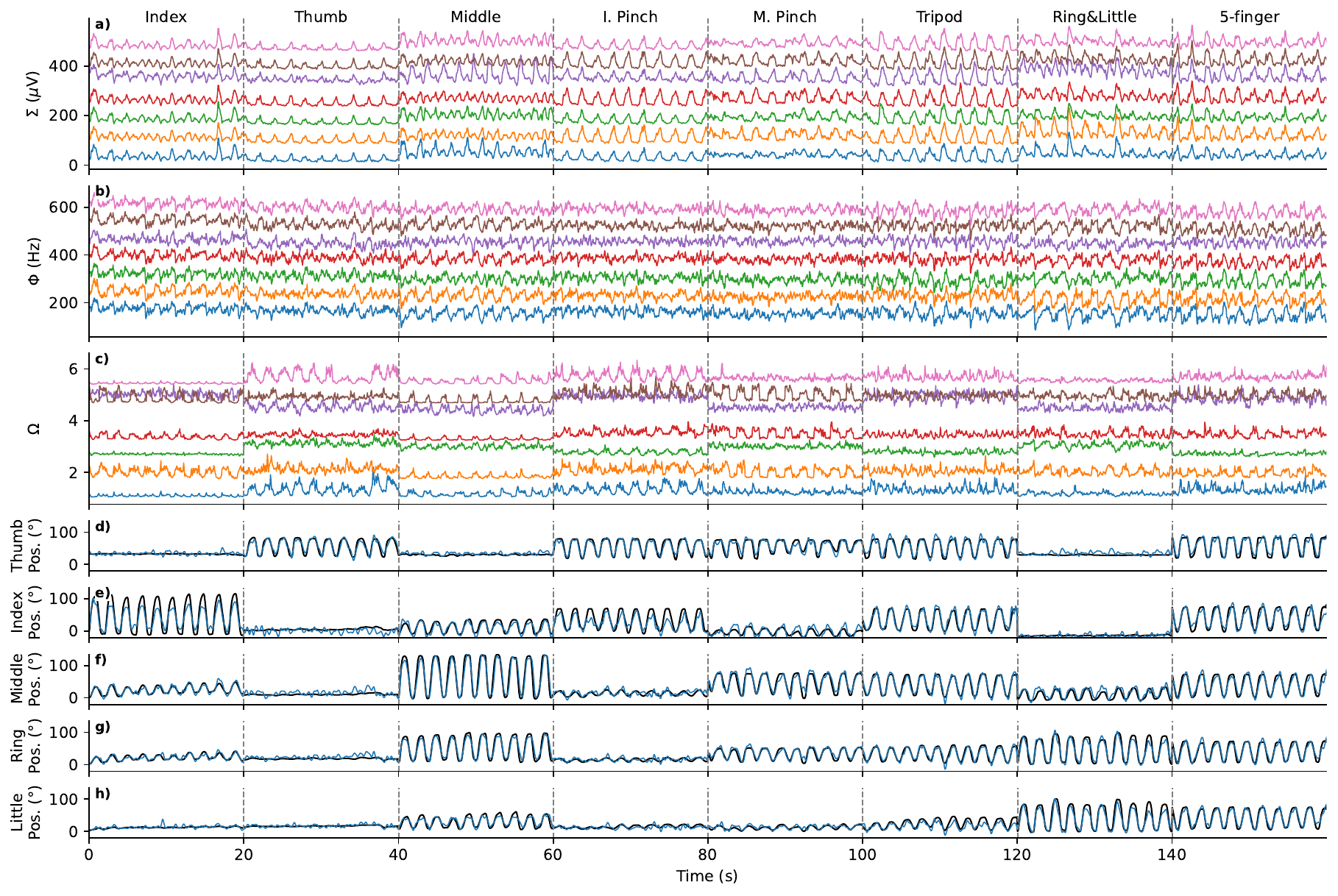}
    \caption{Representative example of the test data, highlighting the feature set and the agreement between computed and predicted finger angular positions. The analysis was performed using the Ridge regression algorithm with a block size of $2 \times 2$, a step of 1, a window duration of 150~ms, and with each sliding window treated as an individual testing sample. Subplots (a)–(c) show the feature set for the seven most relevant blocks, corresponding to effective field strength ($\Sigma$), field-strength variation rate ($\Phi$), and spatial complexity ($\Omega$). Subplots (d)–(h) compare the computed and predicted angular positions for the Thumb, Index, Middle, Ring, and Little fingers.}
    \label{fig:mld_example}
\end{sidewaysfigure*}

\begin{table}[ht]
\centering
\caption{Training time (mean $\pm$ 95\% CI) for each regression algorithm using the multichannel linear descriptors--based block field method (MLD-BFM) feature ($2 \times 2$ blocks, step size 1, 150~ms window, 1 sequence size).}
\label{tab:train_time}
\begin{tabular}{l c}
\toprule
Regression algorithm & Training time (s) \\
\midrule
Ridge & \phantom{0}$0.71 \pm 0.23$ \\
Random Forest (RF) & \phantom{0}$1.85 \pm 0.16$ \\
k-Nearest Neighbors (KNN) & \phantom{0}$1.99 \pm 0.23$ \\
Lasso & \phantom{0}$5.61 \pm 1.08$ \\
Histogram Gradient Boost (HGB) & \phantom{0}$4.31 \pm 0.10$ \\
Multilayer Perceptron (MLP) & $15.52 \pm 0.59$ \\
\bottomrule
\end{tabular}
\begin{flushleft}
\footnotesize
\textit{Note:} All experiments were conducted using the scikit-learn library with CPU-only execution on a machine equipped with an Intel(R) Xeon(R) E-2246G CPU @ 3.60\,GHz, 16\,GB of RAM, running Windows 11 Pro.
\end{flushleft}
\end{table}